\documentclass[intlimits,twoside,a4paper]{article}  

\usepackage{amsmath,amssymb}
\usepackage{graphicx}
\usepackage{siunitx}

\usepackage[T2A]{fontenc}
\usepackage[cp1251]{inputenc}

\usepackage{cmpj2}

\issue{2013}{16}{4}{43006}
\doinumber{10.5488/CMP.16.43006}

\newcommand{\be}{\begin{equation}}
\newcommand{\ee}{\end{equation}}
\newcommand{\bd}{\begin{displaymath}}
\newcommand{\ed}{\end{displaymath}}

\title[Conductivity of aqueous sodium nitrate]{On the influence of molecular structure on  the  conductivity of electrolyte solutions --- sodium nitrate in water\thanks{Dedicated to Prof. Dr. Myroslav Holovko on the occasion of his 70$^\mathrm{th}$ birthday}}

\author[H. Krienke]{H. Krienke\thanks{E-mail: hartmut.krienke@chemie.uni-regensburg.de}}
\address{Universit\"at Regensburg, Institut f\"ur Physikalische und Theoretische Chemie,\\
Universit\"atsstrasse   31,  D-93053 Regensburg, Germany}

\date{Received August 5, 2013, in final form September 8, 2013}
\authorcopyright{H. Krienke, 2013}

\begin{document}

\maketitle

\begin{abstract}
Theoretical calculations  of the conductivity of sodium nitrate in water are presented and compared with experimental measurements. The method of direct correlation force in the framework of the interionic theory is used for the calculation of transport properties in connection with the associative mean spherical approximation (AMSA). The effective interactions between ions in solutions are derived with the help of Monte Carlo and Molecular Dynamics calculations on the Born--Oppenheimer level. This work is based on earlier theoretical and experimental studies of the structure of concentrated aqueous sodium nitrate solutions.

\pacs 05.70.Ln,  61.20.Ja,   61.20.Gy, 61.20.Lc, 77.22-d, 82.45.Gj
\end{abstract}

\section{Introduction}

Over the years a fruitful collaboration with my old friend Myroslav Holovko was devoted to the study of liquids and solutions using the methods of Theoretical Physics, the Statistical Mechanics of many body systems. There are not so many places in the world, where this can be done with the same power and inspiration
as in the Institute of Condensed Matter Physics in Lviv, Ukraine, and an early sign of this kind of research is the famous book by Ihor Yukhnovsky and Myroslav Holovko ``Statistical Theory of Classical Equilibrium Systems'' \cite{Yukh1980} which appeared over 30 years ago in the former Soviet Union (which was perhaps the reason, that it was never translated into English). One of the most interesting goals already in those times was to study the classical systems of charged particles, and the collective coordinate description derived for these systems provided an important basis for the treatment of chemical subjects such as electrolyte solutions. Electrolyte solutions are always around us in everyday life.
They govern biochemical processes in living cells, being essential in many kinds of technological processes.
Studying the different properties of these systems by experimental and theoretical methods is important in order to taylor useful solvents for practical goals \cite{Bart1998}.
Concentrated sodium nitrate solutions in  water  at ambient pressure and 25$^{\circ}$C
belong to the systems for which  very different structural and dynamical properties were derived from measurements. We have recently contributed to these studies
 by  neutron- and x-ray diffraction measurements for a set of four different NaNO$_3$ solutions  \cite{Megy2009}.
 Structural information  gained from the diffraction experiments  is supported by computer simulations and from many body theory. Dynamical properties of electrolyte solutions (diffusion coefficients, conductivity, dielectric relaxation spectra) are also derived from measurements, simulations and theoretical considerations.

 Molecular Dynamics (MD) simulations on Born--Oppenheimer (BO)  level were performed using the DLPOLY--MD code from Daresbury \cite{Smith2002}. We also made Monte Carlo (MC) simulations with
the help of the program system MCFLUID, developed by our group \cite{Fischer2002},  to compare the structural correlation functions with the results of the diffraction measurements. The simulations gave us the spatial distribution functions for pairs of interacting sites of various components of the solution. Various data allowed  us to refine our structural picture and the potential parameters of  our model as well. From the simulation we also derived the values for
the dielectric constants of the solutions, which can be compared with experimental values, e.g., from dielectric relaxation measurements \cite{Bart1995}. The decrease of the dielectric constant of the solution with an increasing solute concentration gives a hint of ion association.

The MD simulations give us also the values for the self-diffusion constants $D_i$, ($i=+,-$) of the ions in the concentrated solutions. These diffusion constants of the ions can be compared with the measured values. They  also allow us to estimate an approximate
value of the specific electrolytic conductivity $\sigma$ via the Nernst--Einstein equation \cite{Bart1998,Hans2006}:
\be\label{e1}
\sigma^{\mathrm{NE}}(c)=\frac{e^2}{k_{\mathrm{B}}T}\left[\rho_{+}z_{+}^2D_{+}(c)+\rho_{-}z_{-}^2D_{-}(c)\right],
\ee
where $k_{\mathrm{B}}$ is the Boltzmann constant, $T$ the temperature, $\rho_i=N_Ac_i$ are the number densities of the ions,  $c_i$ the corresponding concentrations, and $e_i=z_ie$ are the ionic charges.
The equation (\ref{e1}) is correct in the case of infinite diluted solutions ($c_i\rightarrow 0$). On the other hand, new theoretical and experimental studies of ionic liquids, which can be seen as limiting cases of concentrated electrolyte solutions,  show that the Nernst--Einstein  equation is valid for these systems in the case of cross correlations between the moving ions being negligible.
We study its approximate use for our concentrated solutions and compare it with the measured values of the equivalent conductivity of concentrated aqueous sodium nitrate solutions.

 On the other hand, conductivity belongs to the excess properties of ionic solutions with respect to infinite dilution, such as osmotic pressure or mean activity coefficients. These quantities can be calculated by means of Statistical Mechanics from the effective ionic interactions [the ion--ion potentials of mean forces  (pmf) at infinite dilutions] in the framework of the so-called interionic theory.
 According to such a description, the ions are considered to be structured particles interacting by effective (average) forces and moving in a structureless solvent.
 The solvent is solely characterized by macroscopic parameters such as viscosity $\eta$, dielectric constant $\varepsilon_r$ and mobilities $\zeta_i$ of the ions.
This corresponds to a transition from the so-called Born--Oppenheimer (BO) level of the solution, where all molecules and ions are treated on equal footing, to a solvent-averaged or McMillan--Mayer (MM) level, in which only the ions are the subjects of a statistical description \cite{Ebel1988,Bart1998}.

 The introduction of a chemical picture makes it possible to consider the ion association on the MM level. Ion pairs are introduced as a new species, and the interactions between free ions and the pairs have to be specified accordingly.
The reduction to the MM level and to a chemical picture considerably simplifies the statistical mechanical calculations. It allows us to calculate the properties of not excessively concentrated electrolyte systems  starting with the model of charged particles and interacting with essentially weak   Coulomb interactions. However, having a fine structure of the interaction of solvated ions (the ion--ion pmfs in infinite dilution), one has to start with the BO level. We have recently made such a study for the solution of sodium chloride in water-1,4-dioxane mixtures \cite{Krie2013}, and we continue with the consideration of sodium nitrate in water in this article. For this system we have  compared the structural properties derived from x-ray and neutron diffraction measurements with the theoretical predictions from molecular models for solvent and soluted ions  \cite{Megy2009}.

\section{Computer simulations}

The interaction of the molecule sites are modelled with a 12--6--1 Lennard-Jones--Coulomb potential.
For water we used the well known simple
SPC/E model of Berendsen et al. \cite{Bere1987}.
For alkali ions we used, as in the paper \cite{Krie2013}, the model potentials of Palinkas et al. \cite{Pali1977}.
For the nitrate ion we used the values given in \cite{Megy2009}. In addition to the MD simulations for the four solutions studied in \cite{Megy2009} we have also performed MD and MC simulations of  aqueous sodium nitrate solutions with  concentrations of $c=0.65$~mol$\cdot$dm$^{-3}$ \cite{Krie2007} and of $c= 6.67$~mol$\cdot$dm$^{-3}$ \cite{Pete2008}.

Pair correlation functions, energies and other structural quantities were calculated for the  model systems,
and we found a satisfactory agreement with the structural information derived from x-ray and from neutron diffraction measurements for the real solutions \cite{Megy2009}. In addition to the  correlation functions presented in \cite{Megy2009} here we show our results for the ion--ion correlations from MD and MC simulations for six solutions of different concentrations. In the case of molecular ions, we consider every atom of the NO$_3$-ion to be one site with a partial charge.
\begin{figure}[ht]
\includegraphics[width=0.48\textwidth]{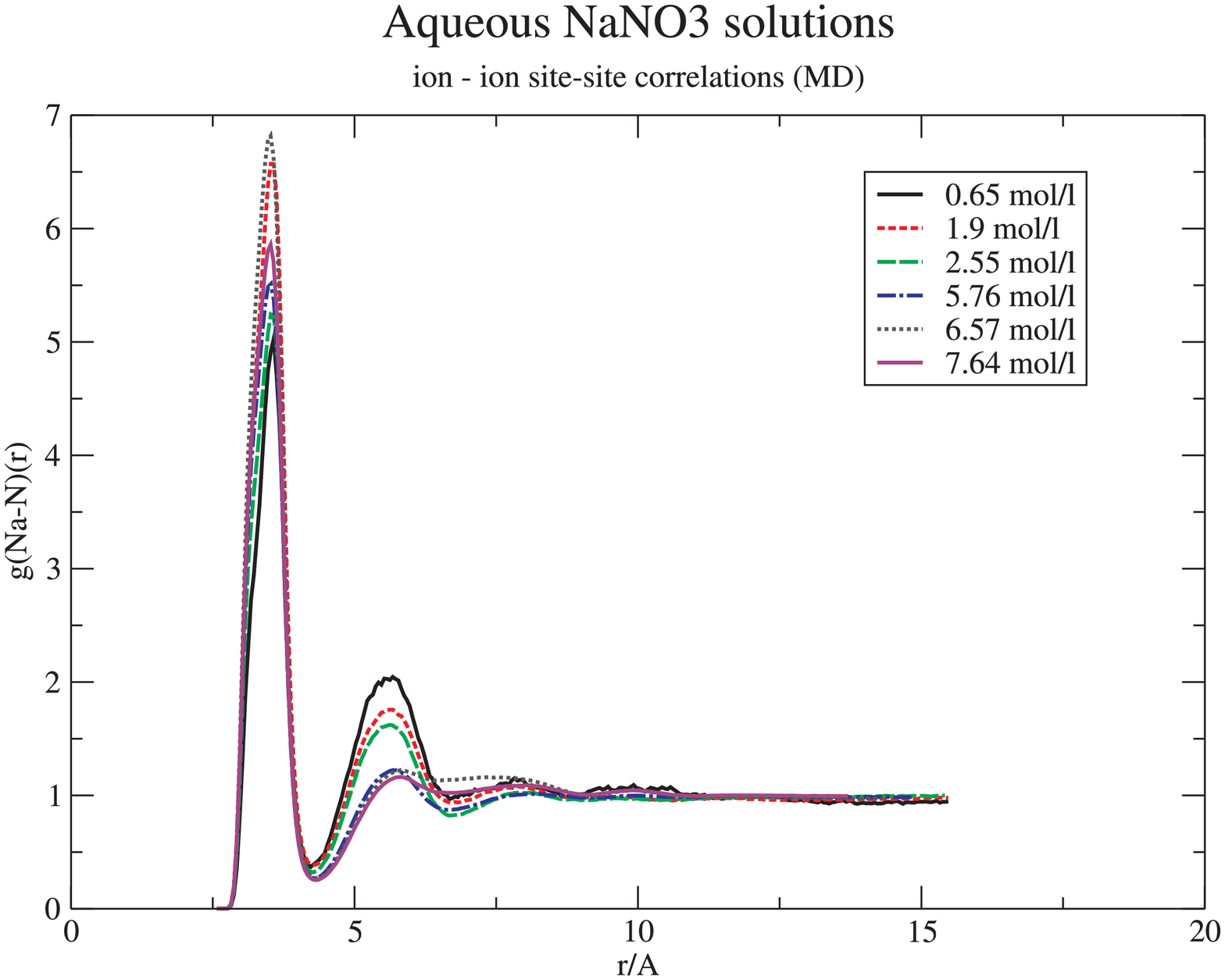}%
\hfill%
\includegraphics[width=0.48\textwidth]{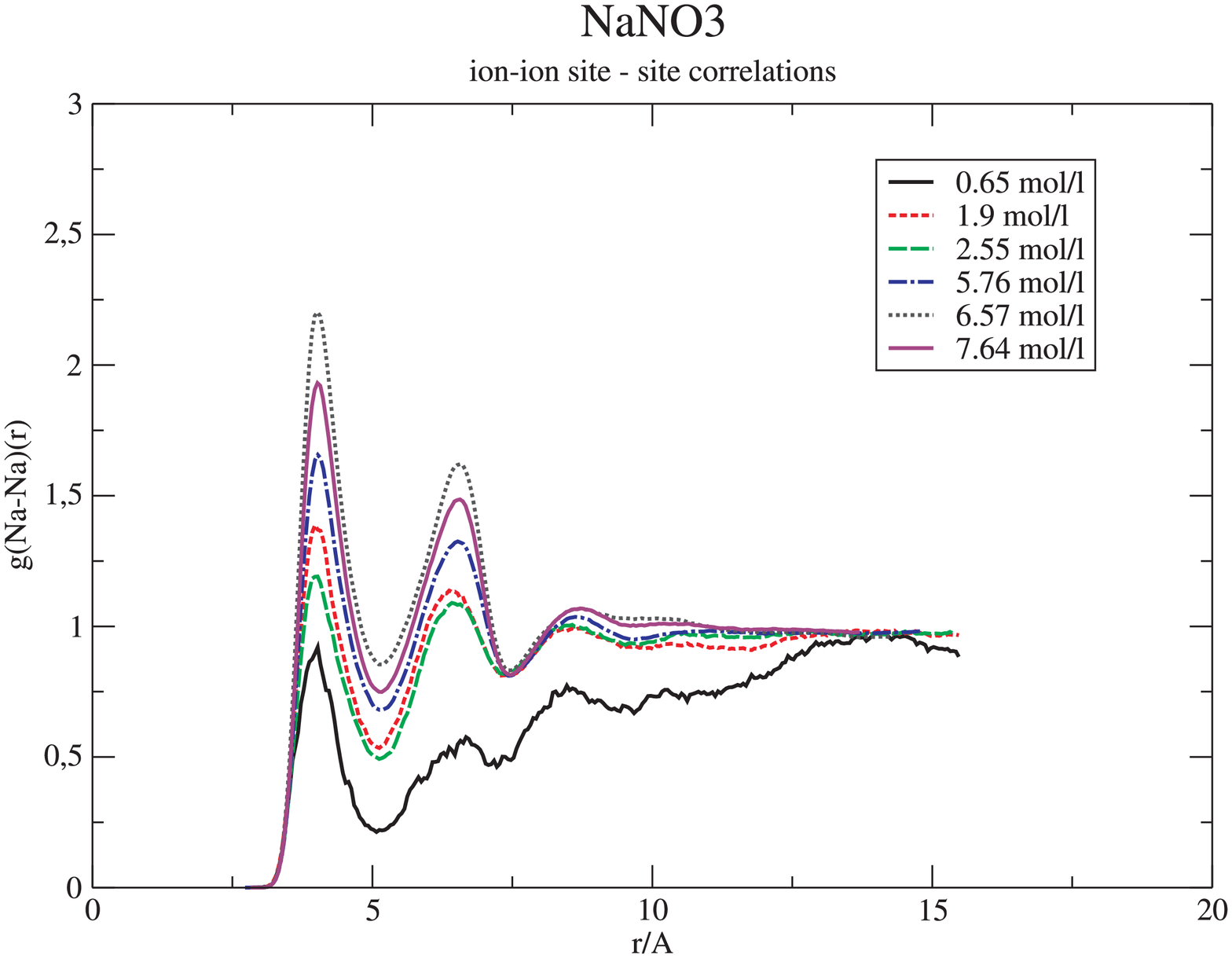}%
\\%
\parbox[t]{0.48\textwidth}{%
\caption{(Color online) Cation--anion pair correlations $g_{\mathrm{Na-N}}(r)$ for six different solutions from MD simulations (see text for a discussion).}
\label{Fig:g1}%
}%
\hfill%
\parbox[t]{0.48\textwidth}{%
\caption{(Color online) Cation--cation pair correlations  $g_{\mathrm{Na-Na}}(r)$  for six different solutions from MD simulations.}
\label{Fig:g2}%
}%
\end{figure}
\begin{figure}[!h]
\centerline{\includegraphics[width=0.48\textwidth]{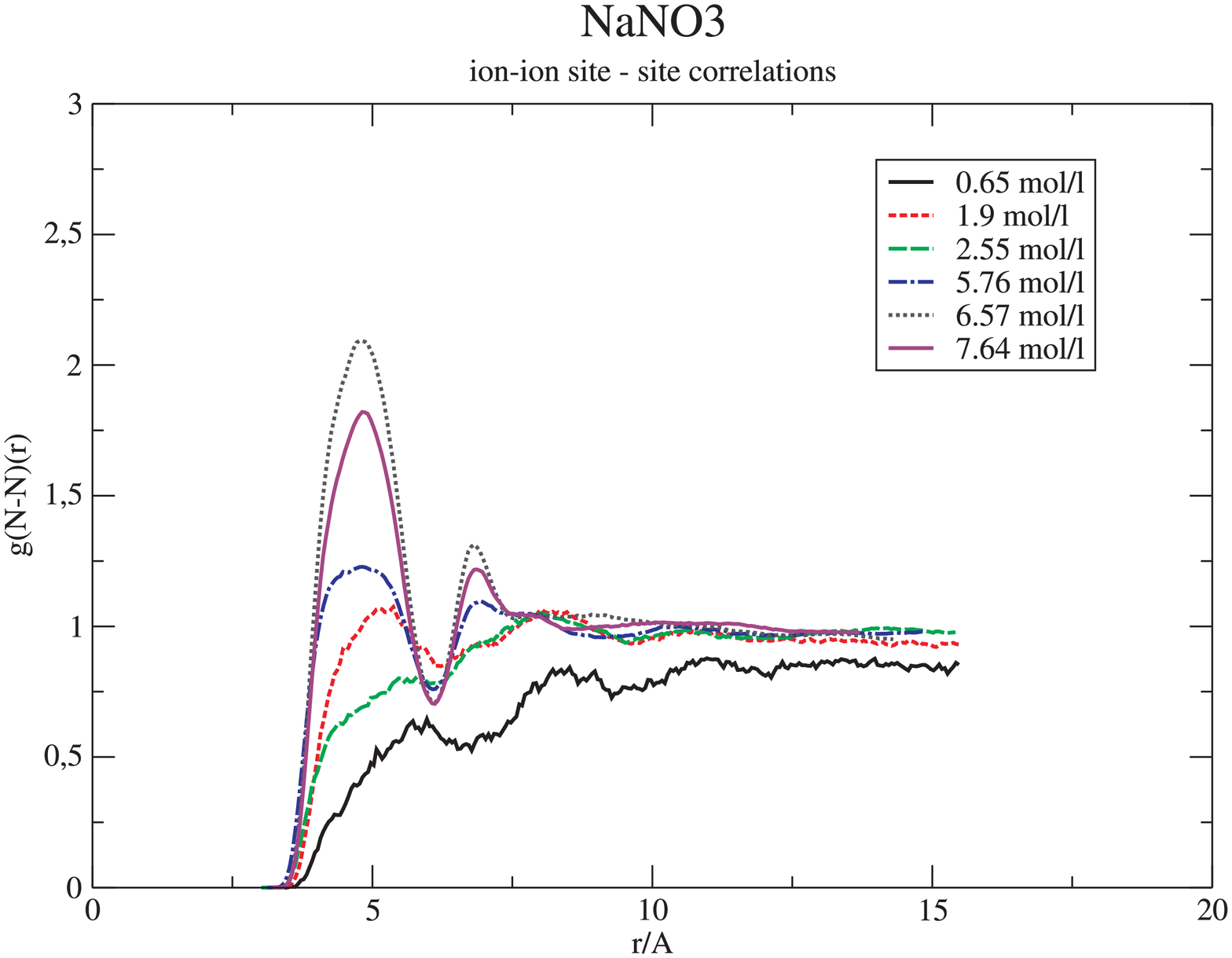}}%
\caption{(Color online) Anion--anion pair correlations  $g_{\mathrm{N-N}}(r)$  for six different solutions from MD simulations.}
\label{Fig:g3}
\end{figure}

 Due to the symmetry of the nitrate potential model, figures~\ref{Fig:g1}--\ref{Fig:g3} also represent  the center--center correlations $g_{ij}(r)$, ($i=+,-$), of the ions in the solutions. From these correlations  the corresponding ion-ion potentials of the mean forces (pmf's) $W_{ij}(r)$ are derived:
\begin{equation}
W_{ij}(r) =-k_{\mathrm{B}}T\ln\left[g_{ij}(r)\right].
\end{equation}

\begin{figure}[th]
\centerline{\includegraphics[width=0.48\textwidth]{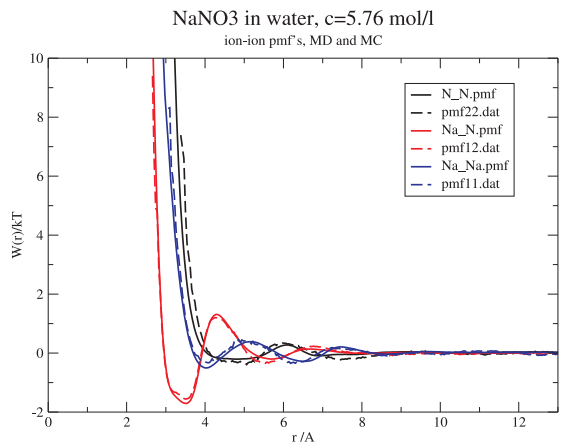}}
\caption{(Color online) Potentials of mean forces between ions in a 5.76 molar aqueous sodium nitrate solution
from MD simulations (full lines) and from MC calculations (broken lines).}
\label{Fig:p1}
\end{figure}

 These functions show a great manifold of structures. This can be seen from  figure~\ref{Fig:p1}, where  the MC
 and MD simulations for the ion--ion pmf's in a
5.76 molar sodium nitrate solution are compared.

 There are  regions of additional attractions as  well as repulsions between the ions, while in the overlap range of the solvation shells specific  interactions  occur, which will be referred to as solvation potentials. However, it is clear that the long range parts of the effective interactions between the ions will be of screened Coulomb- or Debye-type in dilute solutions, i.e., of Coulomb-type in the infinite dilution, with the dielectric constant $ \varepsilon_r$ of the solvent
\begin{displaymath}
\lim_{r \rightarrow \infty,c \rightarrow 0} W_{ij}(r) \rightarrow W^{\infty , \mathrm{Coul}}_{ij}(r) = \frac{z_iz_je^2}{4\pi\varepsilon_0\varepsilon_r r }\,.
\end{displaymath}

The short range parts of the effective ionic interactions result from the overlap of solvation shells. This effect depends only very weakly on the concentration of the solution, as can be seen in figures~\ref{Fig:g1}--\ref{Fig:g3}, which show the ion--ion pair correlation functions for the six solutions studied between 0.65 and 7.5~mol/l. The first maximum in the ($+-$) pair correlation function, and hence the first minimum of the corresponding pmfs, is always in the neighbourhood of 3.5~{\AA}, nearly independent of the concentration of the ions. This will be helpful in constructing simplified  effective ionic interactions for the calculation of conductance in the framework of the solvent averaged McMillan--Mayer theory.

The  mean excess internal energies $U^{\mathrm{ex}}=-E^{\mathrm{C}}$ as well as the solvation energies of cations and anions calculated from the MC simulations are shown in  table~\ref{Tab:t1} for five of the systems studied.
 \begin{table}[ht]
 \caption{Internal excess and solvation energies  in aqueous sodium nitrate solutions (from MC simulations).}
\label{Tab:t1}
\vspace{2ex}
\begin{center}
 \begin{tabular*}{\textwidth} {|l@{\extracolsep{\fill}}|l|l|l|} \hline
System &  Eex[kJ/mol]  & E(Cation.Solv.) & E(Anion.Solv.) \\ \hline \hline
1 ($c=7.64$)  & --132.92  & --169.54 & --167.55\\ \hline
2 ($c=6.67$)  &  --120.54 & --186.46 & --169.85\\ \hline
3 ($c=5.76$)  &  --111.30 & --215.66 & --194.85\\ \hline
4 ($c=2.55$)  &   --75.76 & --313.03 & --307.72\\ \hline
5 ($c=1.9$)   &   --68.16 & --382.24 & --347.22 \\ \hline
\end{tabular*}
\end{center}
 \end{table}

\begin{figure}[ht]
\includegraphics[width=0.48\textwidth]{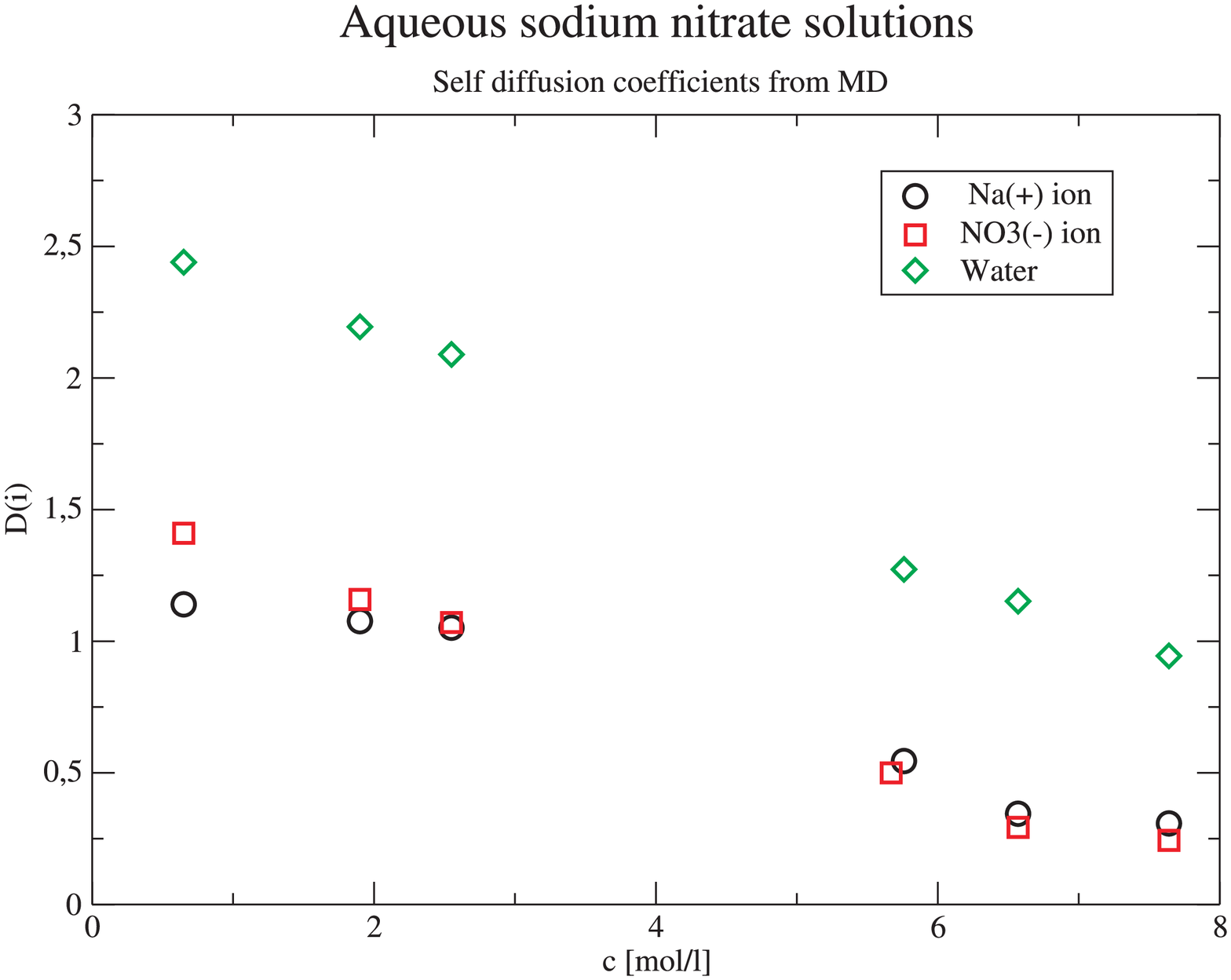}%
\hfill%
\includegraphics[width=0.48\textwidth]{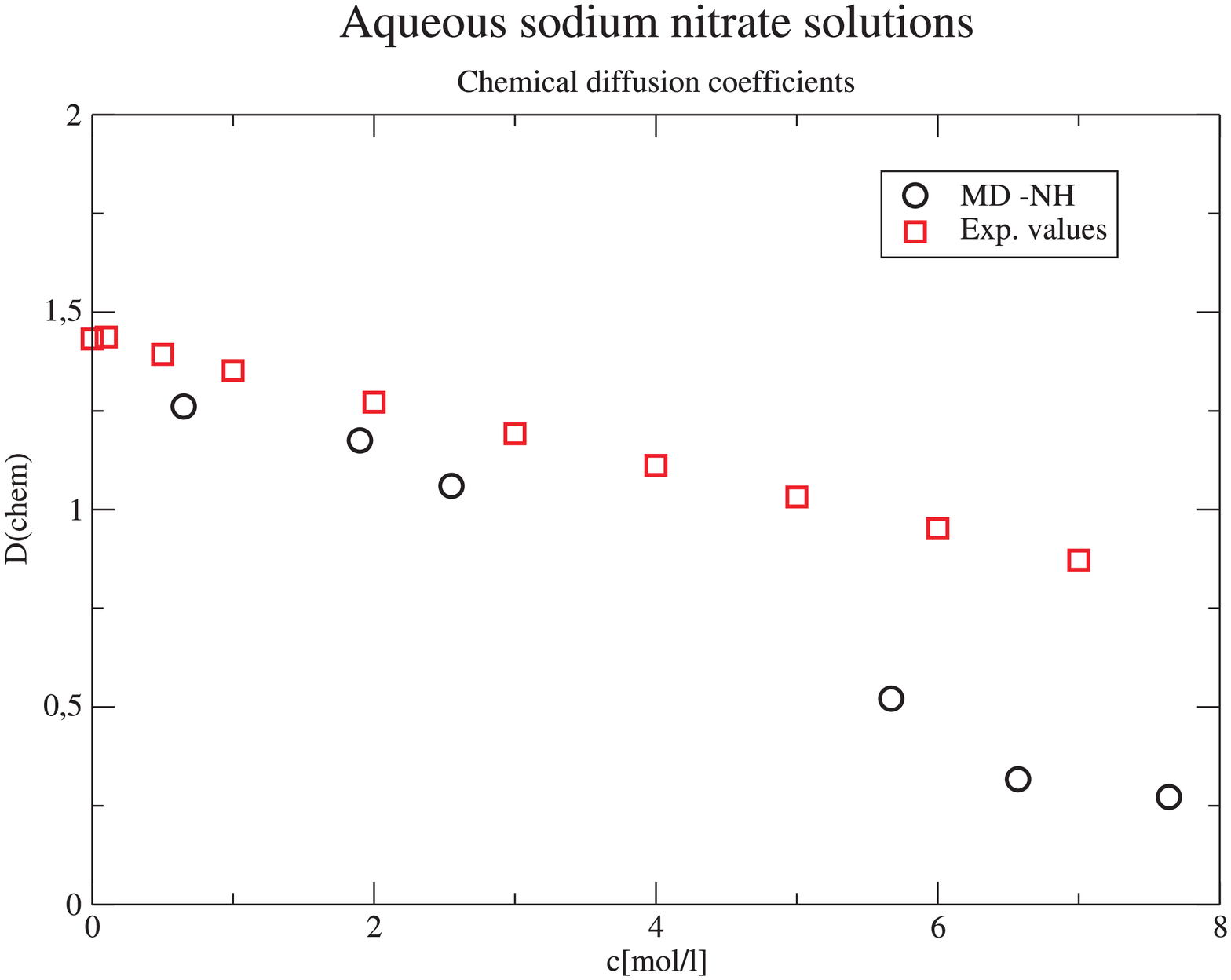}%
\\%
\parbox[t]{0.48\textwidth}{%
\caption{(Color online) Self-diffusion coefficients of  water (diamonds), of Na$^{+}$-ions (circles)  and NO$_3^{-}$-ions  (squares) in the solutions (MD simulations).}
\label{Fig:d1}%
}%
\hfill%
\parbox[t]{0.48\textwidth}{%
\caption{(Color online) Chemical diffusion coefficients  in aqueous sodium nitrate solutions (squares) and Nernst--Hartley  diffusion coefficients from  MD simulations (circles).}
\label{Fig:d2}%
}%
\end{figure}

In figure~\ref{Fig:d1} the self-diffusion coefficients from our MD simulations are shown
for the solvent water and for the sodium and nitrate ions for the six systems studied.

The concentration dependence of the self-diffusion coefficients is very similar to other concentrated aqueous ionic systems \cite{Krie2007}. Using the Nernst--Hartley formula \cite{Bart1998},
\begin{equation}
D_{\mathrm{NH}}^{\infty}= \frac{D_{+}^{\infty}D_{-}^{\infty}(q_{+}^2+q_{-}^2)}
{q_{+}^2D_{+}^{\infty}+q_{-}^2D_{-}^{\infty}}\,, \qquad
q_i^2=\frac{z_i^2e^2}{\varepsilon_0\varepsilon_rk_{\mathrm{B}}T}N_Ac_i\,,
\end{equation}
which is valid for the limiting values of the diffusion coefficients
as an approximate description for the chemical diffusion coefficient D in the concentrated solution,
\begin{equation}
D_{\mathrm{NH}}(c)= \frac{2D_{+}(c)D_{-}(c)}
{D_{+}(c)+D_{-}(c)}\,, \qquad
q_{+}^2=q_{-}^2
\end{equation}
one finds the situation depicted in figure~\ref{Fig:d2}.

The experimental values are from \cite{Janz1970}. As can be seen from the figure, one has the situation, where the $D_{\mathrm{NH}}$ are not a good approximation for the chemical diffusion coefficients at higher concentrations.

\begin{figure}[ht]
\centerline{\includegraphics[width=0.48\textwidth]{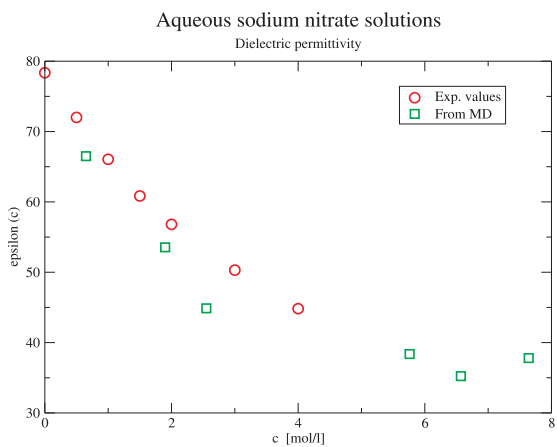}}
\caption{(Color online) Concentration dependent dielectric constants of aqueous sodium nitrate solutions: circles --- experimental values; squares --- from MD simulations.}
\label{Fig:dk1}
\end{figure}

Are there ion pairs in the concentrated sodium nitrate solutions? Our simulations really show decreasing dielectric constants for the higher concentrated solutions, which is in agreement  with the experimental data \cite{Bart1995}. This is depicted in figure~\ref{Fig:dk1}, where the concentration dependent dielectric constants for aqueous sodium nitrate solutions are shown, \cite{Bart1995}, together with our MD simulation values \cite{Krie2007,Pete2008,Megy2009}. Thus, the idea of ion association due to Coulomb forces arises, and a simplified theoretical description in the framework of the interionic theory should consider this aspect.

\begin{figure}[ht]
\centerline{\includegraphics[width=0.48\textwidth]{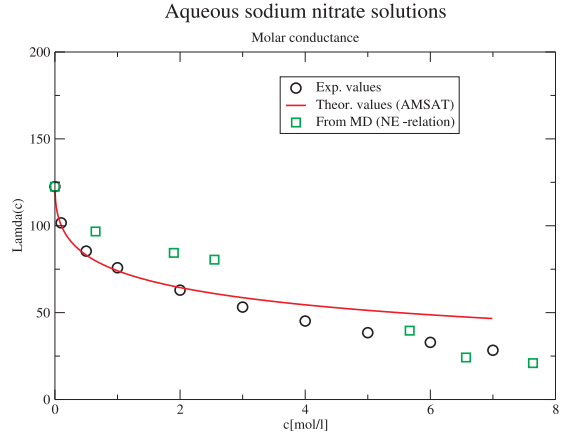}}
\caption{(Color online) Molar conductance of sodium nitrate solution: Circles: experimental values; Line: theoretical calculation from the  AMSA theory for transport with R=3.5 \AA ; Squares: Results from  MD-diffusion coefficients via the Nernst--Einstein equation.}
\label{Fig:l1}       %
\end{figure}

A comparison of  the molar conductance $ \Lambda_{\mathrm{NE}}(c)$ , derived from  the MD self-diffusion coefficients  according to equation~(\ref{e1}) and the relation
\be\label{s54b}
\Lambda_{\mathrm{NE}}(c)=\frac{\sigma_{\mathrm{NE}}(c)}{c}\,,\qquad \Lambda_{\mathrm{NE}}^{\infty}=\lim_{c\rightarrow 0} \Lambda_{\mathrm{NE}}(c)
\ee
with the experimental values of the conductance of the concentrated solutions from \cite{Janz1970} are shown  in figure~\ref{Fig:l1}.

We see, that in the case of very concentrated solutions, a situation similar to the molten salt or ionic liquid case occurs. It has been shown, that the Nernst--Einstein equation (\ref{e1}) is a good approximation for the conductivity of some systems \cite{Padr1991,Rey2006,Yama2010}. The correct conductivity formula is given by \cite{Hans2006}
\be\label{e2}
\sigma^{\mathrm{NE}}(c)=\frac{1}{k_{\mathrm{B}}T}\sum_{i}\rho_{i}z_{i}^2D_{i}(c)[1-\Delta (c)].
\ee

If $\Delta(c)$ is not zero, cross correlations between the ions become important. If the ratio between the experimental value of $\sigma$ and the conductivity calculated from the Nernst--Einstein equation is smaller than unity, then a significant fraction of oppositely charged ions is assumed  to move together
in the time scale of diffusive motion \cite{Rey2006}. $\Delta(c) > 0 $ means, that the movement of an ion pair in the same direction contributes to the self-diffusion, rather than to the electric current \cite{Hans2006}.

With  simplified interaction models for the ions, e.g., with the approximation that
  the $N$-particle potential of
the mean force of the ions at infinite dilution
 $ W_{N}^{\infty}({\bf r}^{N}) $
 is written as a  sum of pair interactions
\be\label{s2a}
W_{N}^{\infty}({\bf r}^{N})=\sum_{i<j}W_{ij}^{\infty}
({\bf r}_{i},{\bf r}_{j})
\ee
 one is able to create a transport theory for electrolytes, which contains elements of statistical mechanics and phenomenological hydrodynamics in the sense of the classical works by Debye, Onsager, Falkenhagen and others, see e.g., \cite{Falk1971,Falk1971a}. One can further assume that ions associate into neutral pairs
\be\label{s60}
 C^{z_{+}}+ A^{z_{-}} \Rightarrow [C^{z_{+}}A^{z_{-}}]^{o}
\ee
and that a mass action law governs the concentration of free ions defining a degree of dissociation $\alpha$.

 The theory has been published in full in the textbook by Barthel et al. \cite{Bart1998}. We have recapitulated it and added some features concerning the effect of ion association on equilibrium and transport properties of electrolytes in systems with low dielectric constants in a recent article \cite{Krie2013}, so we will  recapitulate it here only in a short form in the appendix.

 The impressive feature of this approach is that a simple potential model of charged hard spheres in a continuum
\begin{eqnarray}
 W^{\infty}_{ij}(r) &=& \frac{z_iz_je^2}{4\pi\varepsilon_0\varepsilon_r r }\,, \qquad  \quad r > R,
\\
 W^{\infty}_{ij}(r) &=& \infty\,, \qquad \qquad \qquad r \leqslant  R
\end{eqnarray}
can be used  for the free ions, which gives analytical results for the  transport properties in the framework of the   Mean Spherical Approximation for Transport (MSAT) \cite{Ebel1979,Ebel1981,Bern1992} and of the Associative Mean Spherical Approximation (AMSA) \cite{Ebel1982,Holo1991,Turq1995,Kaly2000,Krie2000,Bart2000,Bart2002}.
The inclusion of a mass action law (MAL) into the transport theory is also briefly discussed in the appendix.

We have made calculations according to the AMSA theory for transport of various aqueous sodium nitrate solutions.
The input parameters are the friction coefficients of the ions in water $\zeta_i$, which also define the single ion conductivities at infinite dilution $\lambda_i^{\infty}$, the viscosity of water $\eta$ and its dielectric constant~$\varepsilon_r$.

Beside these parameters, the  distance of the closest approach $R$ between free ions is the only parameter in this theory. From the diffraction experiments and our simulation results described in figures~\ref{Fig:g1}--\ref{Fig:p1} we see that the value of $ R= 3.50 $ {\AA} is the value of the closest approach for Na$^{+}$- \, and NO$_3^{-}$-ions in our solutions. The  degree of dissociation $\alpha(c)$ is calculated	from these input parameters in the framework of the AMSA theory (see Appendix). The result is shown in figure~9.
\begin{figure}[ht]
\centerline{\includegraphics[width=0.48\textwidth]{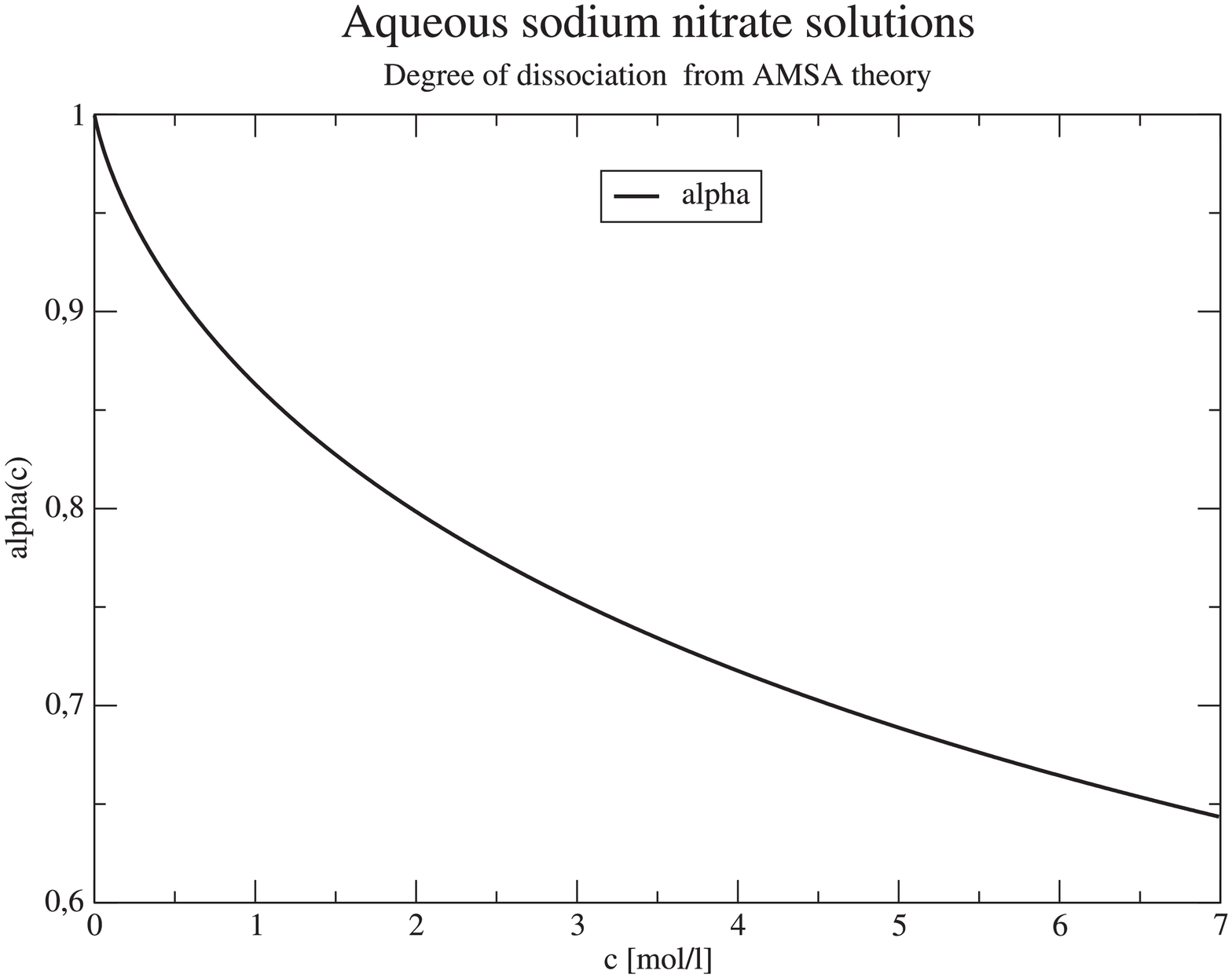}}
\caption{Concentration dependent degree of dissociation of aqueous sodium nitrate solution (from the AMSA theory).}
\label{Fig:al1}
\end{figure}

$\alpha(c)$ decreases in the solutions of higher concentrations. Therefore, the calculation of the conductance  including association is appropriate. Then, we get the molar conductance $ \Lambda $ as
\be\label{s54a}
\Lambda=\alpha\Lambda^{\infty}\left[(1+S_{1}(\alpha)+S_{2}(\alpha)+
S_{3}(\alpha)+S_{4}(\alpha)\right]
\ee
with the $ S $ functions given in \cite{Ebel1981,Ebel1982,Bart1998,Krie2013}.

	With these data, the experimental molar conductivity can be described up to concentrations of several moles per liter, as can be seen in figure~8, where the experimental values from \cite{Janz1970} are compared with the theoretical predictions from the AMSAT theory.

\section{Conclusions}

 The design of special electrolytes for a wide variety of applications is an important goal. Physical and chemical properties of such systems are studied by a variety of experimental and theoretical methods. The combination of simulations, experiment and theory reduces the number of free parameters used for explaining the conductivity data in the solvent. The estimation of the conductance of sodium nitrate in water  is one further example for this kind of studies.

A theory  based on a very simple potential model with only one  parameter (the  distance  $R$, which can be approached by free ions)  is needed in solutions having no excessive concentrations to describe the molar conductivity of aqueous solutions of sodium nitrate. This parameter should be derived from the studies on the Born--Oppenheimer level, where all constituents of the solutions are treated on an equal level. For concentrated solutions, approximative values for the conductance --- at least in our example studied --- are derived from the Molecular Dynamics self-diffusion coefficients of the ions with the help of the Nernst--Einstein equation.

\section*{Acknowledgements}

Most of the MD calculations were started by Emanuel Peter in the course of his diploma work several years ago. His contribution is greatly acknowledged.


\appendix
\section*{Appendix --- conductance on MM level}

When an external force ${\bf F}_{i}^{\mathrm{ext}} $
caused by a homogeneous electric field $ {\bf E} $
\be\label{s12}
{\bf F}_{i}^{\mathrm{ext}}=e_{i}{\bf E}
\ee
acts on ions in a solution, their one-particle flow $ {\bf J}_{i} $ yields Ohm's law for the current density ${\bf j}$:
\be\label{s13}
{\bf j}=\sum_{i}e_{i}{\bf J}_{i}=\sigma {\bf E}.
\ee

The equation for the one-particle flow $
{\bf J}_{i}({\bf r}_{1}) $ with $ \rho_{i}=N_{i}/V $ as the number density
 of species $ i $ is
\be\label{s11}
 {\bf J}_{i}({\bf r}_{1})=\frac{\rho_{i}}{\zeta_{i}}\left[
{\bf F}_{i}^{\mathrm{ext}}({\bf r}_{1})+{\bf F}_{i}^{\mathrm{rel}}({\bf r}_{1})+
{\bf F}_{i}^{\mathrm{el}}({\bf r}_{1})\right]=\frac{\rho_{i}}{\zeta_{i}}
{\bf F}_{i}^{\mathrm{tot}}({\bf r}_{1}).
\ee

 The total force acting on the ions is in the direction of the field ${\bf E}$:
\be\label{s37a}
 {\bf F}_{i}^{\mathrm{tot}}=F_{i}^{\mathrm{tot}}\frac{\bf E}{E}\,, \qquad
F_{i}^{\mathrm{tot}}=e_{i}E+F_{i}^{\mathrm{rel}}+F_{i}^{\mathrm{el}}\,.
\ee

The specific conductivity $ \sigma $ is then expressed as
\be\label{s13a}
\sigma= \sum_{i}\frac{\rho_{i}e_{i}^{2}}{\zeta_{i}}\left[1+
\frac{F_{i}^{\mathrm{rel}}}{e_{i}E}+\frac{F_{i}^{\mathrm{el}}}{e_{i}E}\right].
\ee

The additional forces in equation~\ref{s11} are the relaxation force
$ {\bf F}_{i}^{\mathrm{rel}}({\bf r}_{1}) $ and the electrophoretic force
$ {\bf F}_{i}^{\mathrm{el}}({\bf r}_{1}) $.  $ {\bf F}_{i}^{\mathrm{rel}}({\bf r}_{1}) $ expresses the additional force
acting on an ion of species $ i $ due to the asymmetry of the ion
distribution around it. It is the average of the interaction force acting on ion $ i $ at $
{\bf r}_{1} $ which is obtained  by means of the pair distribution function
$g_{ij}({\bf r}_{1},{\bf r}_{2})$ in the non-equilibrium case:
\be\label{s14}
{\bf F}_{i}^{\mathrm{rel}}({\bf r}_{1})= -\sum_{j}\rho_{j}\int
{\bf \nabla}_{1}W_{ij}^{\infty}({\bf r}_{1},{\bf r}_{2})
g_{ij}({\bf r}_{1},{\bf r}_{2})\rd{\bf r}_{2}\,.
\ee

The  pair distribution function has an equilibrium and a non-equilibrium part:
\be\label{s9a}
g_{ij}({\bf r}_{1},{\bf r}_{2}) =1+h_{ij}({\bf r}_{1},{\bf r}_{2})=
1+h_{ij}^{\mathrm{eq}}(r)+ h_{ij}'({\bf r}_{1},{\bf r}_{2}),
\ee
where  $ h_{ij}'({\bf r}_{1},{\bf r}_{2}) $ corresponds to the
{\it non-equilibrium contribution}
 of the total pair correlation function.

The electrophoretic force may be expressed
 as
\be\label{s14a}
{\bf F}_{i}^{\mathrm{el}}({\bf r}_{1})=
\ \zeta_{i}\sum_{j}\rho_{j}\int {\sf T}_{ij}({\bf r}_{1},{\bf r}_{2})\cdot [e_{j}{\bf E}+
{\bf K}_{ji}({\bf r}_{2},{\bf r}_{1})]  g_{ij}({\bf r}_{1},{\bf r}_{2})]
 \rd{\bf r}_{2}\,.
\ee

The electrophoretic or  hydrodynamic
 interaction tensor ${\sf T}_{ij}({\bf r}_{1},{\bf r}_{2})$ is approximated by its long range part, the Oseen tensor, acting between ions $i$ and $j$ at locations ${\bf r}_{m}$ and
 ${\bf r}_{n}$:
\be\label{s6a}
  {\sf T}_{ij}({\bf r}_{m},{\bf r}_{n})=  \frac{1}{8\pi\eta_{0} r_{mn}}\left({\sf I}_{ij}+
\frac{{\bf r}_{mn}{\bf r}_{mn}}{r_{mn}^2}\right)
 \ee
 ($ {\sf I}_{ij} $ is the unity
tensor,  $ {\bf r}_{mn}={\bf r}_{m} - {\bf r}_{n}$ and ${\bf r}_{mn}{\bf r}_{mn}$ is the dyadic product).

The mean interionic force $ {\bf K}_{ij} $ between an ion of species
 $ i $ at
 $ {\bf r}_{1} $ and an ion $ j $ at $ {\bf r}_{2} $
is given by
\be\label{s16}
{\bf K}_{ij}({\bf r}_{1},{\bf r}_{2})=-
{\bf \nabla}_{1}W_{ij}^{\infty}({\bf r}_{1},{\bf r}_{2})-
\sum_{k}\rho_{k} \int
{\bf \nabla}_{1}W_{ik}^{\infty}({\bf r}_{1},{\bf r}_{3})\frac{g_{ijk}
({\bf r}_{1},{\bf r}_{2},{\bf r}_{3})}{g_{ij}({\bf r}_{1},{\bf r}_{2})}
\rd{\bf r}_{3}\,.
\ee

In the stationary case  ${\mathrm{div}}\, {\bf j}=0 $, Onsagers continuity equation
for the pair distribution functions in the
non-equilibrium situation is   given by the expression \cite{Bart1998}:
\be\label{s26}
{\bf \nabla}_{1}\cdot\tilde{\bf J}_{ji}({\bf r}_{2},{\bf r}_{1})+
{\bf \nabla}_{2}\cdot\tilde{\bf J}_{ij}({\bf r}_{1},{\bf r}_{2})=0.
\ee

The reduced two particle flow
\begin{equation}
\tilde{\bf J}_{ij}({\bf r}_{2},{\bf r}_{1})=
\frac{1}{\zeta_{i}}\left[h_{ij}({\bf r}_{1},{\bf r}_{2})
{\bf F}_{i}^{\mathrm{tot}}({\bf r}_{1})+
{\bf C}_{ij}({\bf r}_{1},{\bf r}_{2})+
\sum_{k}\rho_{k}\int {\bf C}_{ik}({\bf r}_{1},{\bf r}_{3})
h_{kj}({\bf r}_{3},{\bf r}_{2})\rd{\bf r}_{3}
-k_{\mathrm{B}}T{\bf \nabla}_{1}h_{ij}({\bf r}_{1},{\bf r}_{2})\right]
\label{s30}
\end{equation}
contains a  direct correlation force  $ {\bf C}_{ij}({\bf r}_{1},{\bf r}_{2}) $ with an equilibrium
and a non-equilibrium term:
\be\label{s28}
{\bf C}_{ij}({\bf r}_{1},{\bf r}_{2})=
k_{\mathrm{B}}T{\bf \nabla}_{1}c_{ij}^{\mathrm{eq}}({\bf r}_{1},{\bf r}_{2})+
 {\bf C}_{ij}'({\bf r}_{1},{\bf r}_{2}).
\ee

 A corresponding equation is valid for
$ \tilde {\bf J}_{ji}({\bf r}_{2},{\bf r}_{1}) $
\cite{Krem1983}.

 At equilibrium, the flows
$ \tilde{\bf J}_{ij}({\bf r}_{1},{\bf r}_{2}) $ and forces
$ {\bf F}_{i}^{\mathrm{tot}}({\bf r}_{1}) $ vanish and equation~\ref{s30}
 reduces to the usual OZ equation.

 $ {\bf C}_{ij}'({\bf r}_{1},{\bf r}_{2}) $ or, equivalently,  the
non-equilibrium part of the total correlation function,
 $  h_{ij}'({\bf r}_{1},{\bf r}_{2})=h_{ij}({\bf r}_{1},{\bf r}_{2})-h_{ij}^{\mathrm{eq}}({\bf r}_{1},{\bf r}_{2}) $  must now be calculated.

For binary electrolytes, ($ i,j=+,- $),
the symmetry relations  in the linear approximations in $\bf E$
lead to ${\bf C}_{ii}'({\bf r})={\bf 0} $ and $
 h_{ii}'({\bf r})=0 $. Also $
{\bf C}_{-+}'({\bf r},{\bf E})=
-{\bf C}_{+-}'(-{\bf r},{\bf E})$
and
$ h_{-+}'({\bf r},{\bf E})=h_{+-}'(-{\bf r},{\bf E})$.

After a Fourier transformation according to
\be\label{s38a}
\hat{f}({\bf q})=\int \exp(i{\bf q}\cdot{\bf r})f(\bf r)\rd{\bf r}
\ee
of the continuity equation,
 a relation for $\hat{h}_{+-}'({\bf q})= \hat{h}_{+-}'(q,\theta') $ results in the ${\bf q}$-space
\begin{eqnarray}
\hat{h}_{+-}'(q,\theta')&=&\frac{-\ri(w_{+-}F_{+}^{\mathrm{tot}}-w_{-+}F_{-}^{\mathrm{tot}})
\hat{h}_{+-}^{\mathrm{eq}}(q)}{k_{\mathrm{B}}Tq[1-\rho_{+}w_{+-}\hat{c}_{++}^{\mathrm{eq}}(q)
-\rho_{-}w_{-+}\hat{c}_{--}^{\mathrm{eq}}(q)]}\cos \theta'
\nonumber\\
&&{}
-\frac{ \ri[1+ \rho_{+}w_{-+}\hat{h}_{++}^{\mathrm{eq}}(q)+
\rho_{-}w_{+-}\hat{h}_{--}^{\mathrm{eq}}(q)]}{k_{\mathrm{B}}Tq^2[1-\rho_{+}w_{+-}
\hat{c}_{++}^{\mathrm{eq}}(q)
-\rho_{-}w_{-+}\hat{c}_{--}^{\mathrm{eq}}(q)]}
{\bf q}\cdot\hat{\bf C}_{+-}'({\bf q}).
\label{s39}
\end{eqnarray}

To derive equation~(\ref{s39}), one has to consider the relations between the vectors ${\bf E}$,  ${\bf r}$ and  ${\bf q}$. One has
\be\label{s42}
\frac{{\bf r}\cdot{\bf E}}{rE}=\cos\theta.
\ee

Then, the non-equilibrium term  $ h_{ij}'(r,\theta) $
of the total correlation function can be formulated as
\be\label{s43}
h_{ij}'(r,\theta)=\mu_{ij}y(r)\cos\theta, \qquad
\mu_{ij}=\frac{1}{k_{\mathrm{B}}T}\left(w_{ij}F_{i}^{\mathrm{tot}}-w_{ji}F_{j}^{\mathrm{tot}}\right)
\ee
with the radial part $ y(r) $ and $w_{ij}$ defined as
\be\label{s43aa}
w_{ij}=\frac{1}{\zeta_{i}}\left[\frac{1}{\zeta_{i}}+
\frac{1}{\zeta_{j}}\right]^{-1}
,\qquad
w_{ji}=\frac{1}{\zeta_{j}}\left[\frac{1}{\zeta_{i}}+
\frac{1}{\zeta_{j}}\right]^{-1}\,.
\ee

The three-dimensional Fourier transform of
$ h_{ij}'({\bf r}) $
is denoted by $ \hat{h}_{ij}'({\bf q}) $ and given by
\be\label{s44b}
 \hat{h}_{ij}'({\bf q})=\hat{h}_{ij}'(q,\theta')=\ri q\mu_{ij}\omega(q)\cos \theta'
\ee
with the help of function  $\omega(q)$ containing the $q$-dependent part of the transformed nonequilibrium pair correlation function  $ \hat{h}_{ij}'({\bf q})$. The angle $\theta'$ in ${\bf q}$-space  is defined by
\be\label{s42a}
\frac{{\bf q}\cdot{\bf E}}{qE}=\cos\theta'.
\ee

Due to the angular dependence expressed in equations~(\ref{s43}) and
(\ref{s42a}), one finds for the function $ \omega(q) $ defined in equation~(\ref{s44b})
\be\label{s45}
\omega(q)=-4\pi \frac{1}{q}\frac{\rd}{\rd q}\left[\int_{0}^{\infty}
\frac{\sin qr}{qr}y(r)r \rd r\right].
\ee

The back transformation of this function yields for the radial part $y(r)$ of $ h_{ij}'(r,\theta) $,  equation~(\ref{s43}):
\be\label{s46}
y(r)=\frac{\rd}{\rd r}\left[\frac{1}{2\pi^{2}}\int_{0}^{\infty}
\frac{\sin qr}{qr}\omega(q)q^2 \rd q\right]=\frac{\rd}{\rd r}\omega(r).
\ee

As in the case of the equilibrium OZ equation,
the second relation between $ h_{ij} $ and $ {\bf C}_{ij} $
is needed to calculate the correlation functions.
A non-equilibrium analogue for the HNC closure is \cite{Krem1983,Ebel1988}
\begin{equation}
{\bf C}_{ij}({\bf r}_{1},{\bf r}_{2})=
\left[-{\bf \nabla}_{1}W_{ij}^{\infty}({\bf r}_{1},{\bf r}_{2})\right]
g_{ij}({\bf r}_{1},{\bf r}_{2})+
h_{ij}({\bf r}_{1},{\bf r}_{2})\sum_{k}\rho_{k}\int
{\bf C}_{ik}({\bf r}_{1},{\bf r}_{3})
h_{kj}({\bf r}_{3},{\bf r}_{2})\rd{\bf r}_{3}\,.
\label{s31}
\end{equation}

A simplified closure relation for the nonequilibrium part of the direct correlation force is
\be\label{s47}
{\bf C}_{+-}'({\bf r})=-h_{+-}'({\bf r})
{\bf \nabla}\left\{W_{+-}^{\infty}(r)-k_{\mathrm{B}}T[h_{+-}^{\mathrm{eq}}(r)-c_{+-}^{\mathrm{eq}}(r)]\right\}.
\ee

The system of equations equation~(\ref{s39}) and  equation~(\ref{s31}) or  equation~(\ref{s47}) is solved by numerical methods for general ionic interactions $ W_{ij}^{\infty}(r)$.

If the interaction  is represented by that of charged hard spheres of equal diameter $ R $, a natural boundary condition for the relative flows arises  at the contact distance  $R$
\be\label{s48c}
({\bf r}_{1}-{\bf r}_{2})\cdot [ \tilde{\bf J}_{ij}({\bf r}_{1},{\bf r}_{2}) -
\tilde{\bf J}_{ji}
({\bf r}_{2},{\bf r}_{1})]=0 \quad \text{if} \quad
 |{\bf r}_{1}-{\bf r}_{2}|=R.
\ee

Describing the equilibrium parts of the direct and total correlation functions in the Mean Spherical Approximation, one finds for the nonequilibrium contributions in the corresponding MSA conditions~\cite{Krem1983}:
\be\label{s47a}
{\bf C}_{+-}'({\bf r})={\bf 0}
\ee
and
\be\label{s47b}
h_{+-}'({\bf r})=0 , \qquad  r\leqslant  R.
\ee

Defining in the following equations  the Debye parameter $ \kappa$ and the Bjerrum parameter $ b$:
\be\label{s58b}
\kappa^{2}=\frac{|z_{+}z_{-}|e^{2}(\rho_{+}+\rho_{-})}{\varepsilon_0\varepsilon_rk_{\mathrm{B}}T} \,, \qquad  b=\frac{|z_{+}z_{-}|e^{2}}{4\pi\varepsilon_0\varepsilon_rk_{\mathrm{B}}TR}
\ee
in the framework of the Mean Spherical
Approximation for Transport  (MSAT)  for the help function $ \omega(r)$ one finds  the result which assures the flow condition, equation~(\ref{s48c}) \cite{Ebel1979,Ebel1981,Ebel1982}:
\be\label{s53}
\omega(r) = \frac{R^{3}}{2r}\left(b+\frac{1}{3}\right)\exp(-\kappa r)-
\frac{1}{2\pi^{2}}\int_{0}^{\infty}\frac{q}{r}
\frac{\sin(qr)h_{+-}^{\mathrm{eq}\mathrm{,MSA}}(q)}{(q^{2}+0.5\kappa^{2})}\rd q
\ee
$ h_{+-}^{\mathrm{eq}\mathrm{,MSA}}(q) $ is the Fourier transform of the equilibrium
total correlation function of charged hard spheres  in the MSA. With these results, the nonequilibrium pair correlation function $ {h_{+-}^{\mathrm{MSA}}}'(r,\theta) $ is calculated as well as the relaxation and electrophoretic  contributions to the conductivity.

The results for the MSAT conductance derived in \cite{Bern1992,Turq1995} are in agreement with the results by Ebeling and coworkers \cite{Ebel1979,Ebel1981,Ebel1982}.

 If it  is assumed
that the cations and anions  form ion pairs,
  the  density of the ionic species $i $ splits into two parts,
namely  the density of free ions
$ \alpha \rho_{i} $, and the density of ion pairs $(1-\alpha)
\rho_{i} $ \cite{Holo1991,Kaly2000,Krie2000,Bart2000,Bart2002}
. They are connected by a mass action law (MAL) with
Ebeling's  mass action constant $K_{\mathrm{A}}^{\mathrm{E}}$ \cite{Ebel1968,Falk1971}
for ion association, which depends only on the ionic charges and the distance of the closest approach $R$ --- the ion diameter in the RPM:
\be\label{s61}
\frac{1-\alpha}{\alpha^{2}}=cK_{\mathrm{A}}^{\mathrm{E}}\frac{(y_{\pm}')^{2}}
{y_0'} , \qquad
K_{\mathrm{A}}^{\mathrm{E}}=
8\pi N_{\mathrm{A}} R^{3}\sum_{m=2}^{\infty}\frac{b^{2m}}{(2m)!(2m-3)}\,.
\ee

In dilute solutions, the mean activity coefficients of the free ions,
 $ y_{\pm}' $ , and the activity coefficients of the ion pairs,  $ y_{0}'$,
are substituted by their electrostatic contributions
 $ y_{\pm}^{\mathrm{el'}} $ and to $ y_{0}^{\mathrm{el'}} $. In the framework of the AMSA these quantities are given by
\be\label{s62}
\ln(y_{\pm}^{\mathrm{el'}})=-b\frac{\Gamma_{\mathrm{B}} R}
{(1+\Gamma_{\mathrm{B}} R)} , \qquad  \ln(y_{0}^{\mathrm{el'}})=-b\frac{(\Gamma_{\mathrm{B}} R)^2}
{(1+\Gamma_{\mathrm{B}} R)^2}\,.
\ee

In these equations, a new
screening parameter $ \Gamma_{\mathrm{B}} $ appears, which is connected with the Debye $\kappa$  and the degree of dissociation $\alpha$ \cite{Bern1996}:
\be\label{s63}
4\Gamma_{\mathrm{B}}^{2}(1+\Gamma_{\mathrm{B}}R)^{2}=
\kappa^{2}\frac{(\alpha+\Gamma_{\mathrm{B}}R)}
{(1+\Gamma_{\mathrm{B}}R)}\,.
\ee

Equations~(\ref{s61}) to (\ref{s63}) are now combined to calculate the concentration dependent degree of dissociation $\alpha (c)$ for the system studied. For $R=3.5$~\AA, its value is given in figure~\ref{Fig:al1}.

\ukrainianpart

\title{Про вплив молекулярної структури на провідність розчинів електролітів --- нітрат натрію у воді}

\author{Г. Крінке}

\address{Університет Регензбурга, Інститут фізики і теоретичної хімії,
  D--93053 Регензбург, Німеччина}

\makeukrtitle
\begin{abstract}
Представлено теоретичні обчислення провідності нітрату натрію у воді, порівняні з експериментальними вимірюваннями. Метод
прямої кореляційної сили в рамках міжіонної теорії використано для
обчислення властивостей транспорту у зв'язку з асоціативним
середньосферичним наближенням (AMSA). Ефективні взаємодії між іонами
в розчинах отримані з обчислень методами Монте Карло і
молекулярної динаміки на рівні Борна-Оппенгеймера. Ця робота
базується на попередніх теоретичних і експериментальних досліженнях
структури концентрованих водних розчинів нітрату натрію.

\keywords електроліти, водний розчин нітрату натрію, структурні і
транспортні властивості, провідність

\end{abstract}

\end{document}